\documentclass{elsart5p}

 \usepackage{epsfig}

\usepackage{amssymb}

\begin{document}

\begin{frontmatter}



\title{Background reduction and sensitivity for germanium double beta decay experiments}


\author{H. G\'{o}mez\corauthref{cor1}}, S. Cebri\'{a}n, J. Morales, J.A. Villar

\address{Laboratorio de F\'{i}sica Nuclear y Altas Energ\'{i}as, Universidad de Zaragoza, 50009 Zaragoza, Spain}

\corauth[cor1]{Corresponding  Author: Laboratorio de F\'{i}sica
Nuclear y Altas Energ\'{i}as, Facultad de Ciencias, Pedro Cerbuna
12, 50009 Zaragoza, Spain. Phone number: 34 976761246. Fax number:
34 976761247}

\ead{hgomez@unizar.es}

\begin{abstract}
Germanium detectors have very good capabilities for the
investigation of rare phenomena like the neutrinoless double beta
decay. Rejection of the background entangling the expected signal
is one primary goal in this kind of experiments. Here, the
attainable background reduction in the energy region where the
neutrinoless double beta decay signal of $^{76}$Ge is expected to
appear has been evaluated for experiments using germanium
detectors, taking into consideration different strategies like the
granularity of the detector system, the segmentation of each
individual germanium detector and the application of Pulse Shape
Analysis techniques to discriminate signal from background events.
Detection efficiency to the signal is affected by background
rejection techniques, and therefore it has been estimated for each
of the background rejection scenarios considered. Finally,
conditions regarding crystal mass, radiopurity, exposure to cosmic
rays, shielding and rejection capabilities are discussed with the
aim to achieve a background level of $\sim$10$^{-3}$ c keV$^{-1}$
kg$^{-1}$ y$^{-1}$ in the region of interest, which would allow to
explore neutrino effective masses around $\sim$40 meV.
\end{abstract}

\end{frontmatter}


\section{Introduction}
\label{Intro}

Investigation of neutrinoless Double Beta Decay (DBD) can shed
light on interesting pending questions like the absolute values of
the neutrino mass and the properties of neutrinos under
CP-conjugation \cite{reviewsdbd}. Germanium detectors enriched in
double beta decay emitter $^{76}$Ge, offer important advantages
for this investigation in comparison with other kind of detectors
and nuclei \cite{reviewsmerits,avignonenjp}: excellent energy
resolution, high purity materials and powerful background
rejection capabilities, well established detector technologies,
favorable nuclear matrix element, high transition energy Q around
2039 keV \cite{q}, \dots. Indeed, germanium double beta decay
experiments (Heidelberg-Moscow and IGEX) have provided the most
restrictive bounds on the effective neutrinos mass
\cite{hm,igex2002}.

Both Heidelberg-Moscow and IGEX experiments, presently finished,
followed similar strategies using several massive high-purity
germanium detectors (HPGe) (around 2 kg each) with close-end
coaxial geometries together with active and passive shieldings in
deep underground laboratories. As a continuation of these
experiments, next-generation projects as Majorana \cite{majorana}
and GERDA \cite{gerda}, have been proposed incorporating different
innovations like using segmented germanium detectors or operating
naked crystals in cryogenic liquids, like nitrogen or argon.

Important advances in high-purity germanium detector technologies
have been achieved in the last years \cite{advances}, allowing the
construction of large efficiency HPGe crystals and developing the
monolithic segmentation technique which provides both interaction
position and energy information. The analysis of pulse shapes in
highly segmented germanium detectors (taking into account not only
net signals in a segment but also induced transient signals in the
neighboring segments) has revealed as a promising technique for
three-dimensional position determination \cite{kroll}. First
results of the operation of segmented germanium crystals within
DBD projects have recently been presented
\cite{majoranasegmented,gerdasegmented}. Important achievements
have been obtained in other contexts for the spatial resolution of
events (see for instance Refs.
\cite{niedermayr}-\cite{genetic}), greatly surpassing the position
determination capabilities of previous DBD germanium experiments
based on pulse shape discrimination in conventional germanium
detectors \cite{gonzalez}. The main difference between real double
beta decay and background events is that the former leave generally
only one energy deposit inside the crystal ("monosite" events),
whereas the latter leave in many cases more than one ("multisite"
events). For this reason, thanks to the analysis of pulse shapes,
only background events producing one energy deposit in the crystal
can be mistaken with real double beta decay events. One goal of this
work is therefore to evaluate the background reduction attainable at
present for germanium detectors in the region between 2 and 2.1 MeV
where the neutrinoless double beta decay signal of $^{76}$Ge is
expected to appear. Reduction of background is based on the
granularity of the detector system, on the segmentation of each
individual germanium detector and on the application of Pulse Shape
Analysis (PSA) techniques to discriminate signal from background
events.

In underground experiments searching for rare events like the
nuclear double beta decay, the background entangling the expected
signal comes mainly from environmental gamma radiations (including
radon emissions) and neutrons at the laboratory, radioactive
impurities (either primordial or cosmogenically induced) in the
materials of the experimental set-up and cosmic muons arriving
even deep underground \cite{heusser,formaggio}. To explore
effective neutrino masses around 40 meV, a background level
$\sim$10$^{-3}$ c keV$^{-1}$ kg$^{-1}$ y$^{-1}$ in the region of
interest must be achieved in germanium double beta decay
experiments. The two main sources of the background registered in
the region of interest for the neutrinoless double beta decay of
$^{76}$Ge have been identified in previous experiments like IGEX
to be cosmogenic activation of germanium detectors (mainly
$^{68}$Ge and $^{60}$Co) \cite{aalseth} and external gamma
background above 2 MeV coming from $^{232}$Th and $^{238}$U
chains. In fact, in present projects of new experiments these two
sources continue to be a dominant background \cite{majorana,
gerda}. Therefore, in this work the particular background
components thought to be the most significant ones will be taken
into account: cosmogenic $^{68}$Ge and $^{60}$Co produced in the
germanium crystal and the 2614.5 keV emission from $^{208}$Tl in
the $^{232}$Th chain. Precise conditions necessary to achieve a
background level $\sim$10$^{-3}$ c keV$^{-1}$ kg$^{-1}$ y$^{-1}$
in the region of interest from these sources will be discussed.
Recently, a Monte Carlo study of the background achievable in the
GERDA experiment by anti-coincidence cuts between crystals and
segments has been published considering the main radioactive
impurities in the set-up \cite{gerdabackground}; it has also been
shown in the context of this experiment that muon-induced
contribution to background can be of $\sim$10$^{-4}$ c keV$^{-1}$
kg$^{-1}$ y$^{-1}$ provided that muon veto system is used
\cite{gerdamuons}.

The paper is organized in the following way. In Sec. \ref{raw} the
raw backgrounds expected from the relevant sources are studied,
including results from a recent evaluation of the production rates
of the most relevant cosmogenic products in germanium detectors at
sea level. Then, the analysis of the effects of each one of the
three reduction strategies considered (granularity, segmentation
and PSA) for the different background sources will be presented in
Sec. \ref{rejection}. Detection efficiency to the neutrinoless DBD
signal and corresponding sensitivity will be studied in Sec.
\ref{sensitivity} assuming the different background rejection
scenarios. Finally, results and conclusions will be discussed in
Sec. \ref{conclusions}.

\section{Raw backgrounds} \label{raw}

An attempt has been made to estimate the raw contributions to the
detector counting rates coming from the main sources of background
taken into account, before applying any technique for background
reduction.

For this estimate, and for the study of the different background
reduction strategies presented in the next section, a set of Monte
Carlo simulations using GEANT4 \cite{geant4} package has been
developed. As the first goal was focused on the crystal geometry,
a natural germanium crystal without shielding has been defined as
detector. Internal contaminations of the crystal are emitted
homogeneously inside it, whereas for external sources, the
corresponding gammas were emitted homogeneously and isotropically
from the surface of an external sphere. Standard GEANT4 models,
including those specific for low energy, have been used for all
the processes, isotopes decays, and particles simulated. For every
simulation made, position and energy of each interaction produced
in an event have been registered. This information has allowed us
to make different analysis of the obtained data in the Region of
Interest (RoI) between 2 and 2.1 MeV. It has been simulated a
number of events big enough to obtain a negligible statistical
error (below 2\% for all the studies made).

\subsection{Cosmogenic radioactivity} \label{cosmogenics}

Long-lived radioactive nuclei induced by the exposure of the
materials of the set-up to cosmic rays at sea level (during
fabrication, transport and storage) may become very problematic for
rare event experiments which operate in deep underground locations,
using active and passive shields and selecting carefully radiopure
components. Therefore, materials must be kept shielded against the
hadronic component of the cosmic rays to prevent cosmogenic
activation, flying must be avoided and the exposure on the surface
of the Earth should be reduced as much as possible. Since these
requirements usually complicate the preparation of experiments (for
example, crystal growth and mounting of detectors) it would be
desirable to have reliable tools to quantify the real danger of
exposing the materials to cosmic rays.

For these reasons, cosmogenic activation in germanium double beta
decay experiments was specifically studied in Ref.
\cite{cosmogenicstaup2005}. Using these and previous results from
the literature, together with specific simulations of the response
to the cosmogenic background of germanium detectors, the expected
counting rate due to this effect in the neutrinoless DBD region
has been evaluated.

\subsubsection{Production rates} \label{Production Rates}

\begin{figure}
\centering
  \includegraphics[height=.2\textheight]{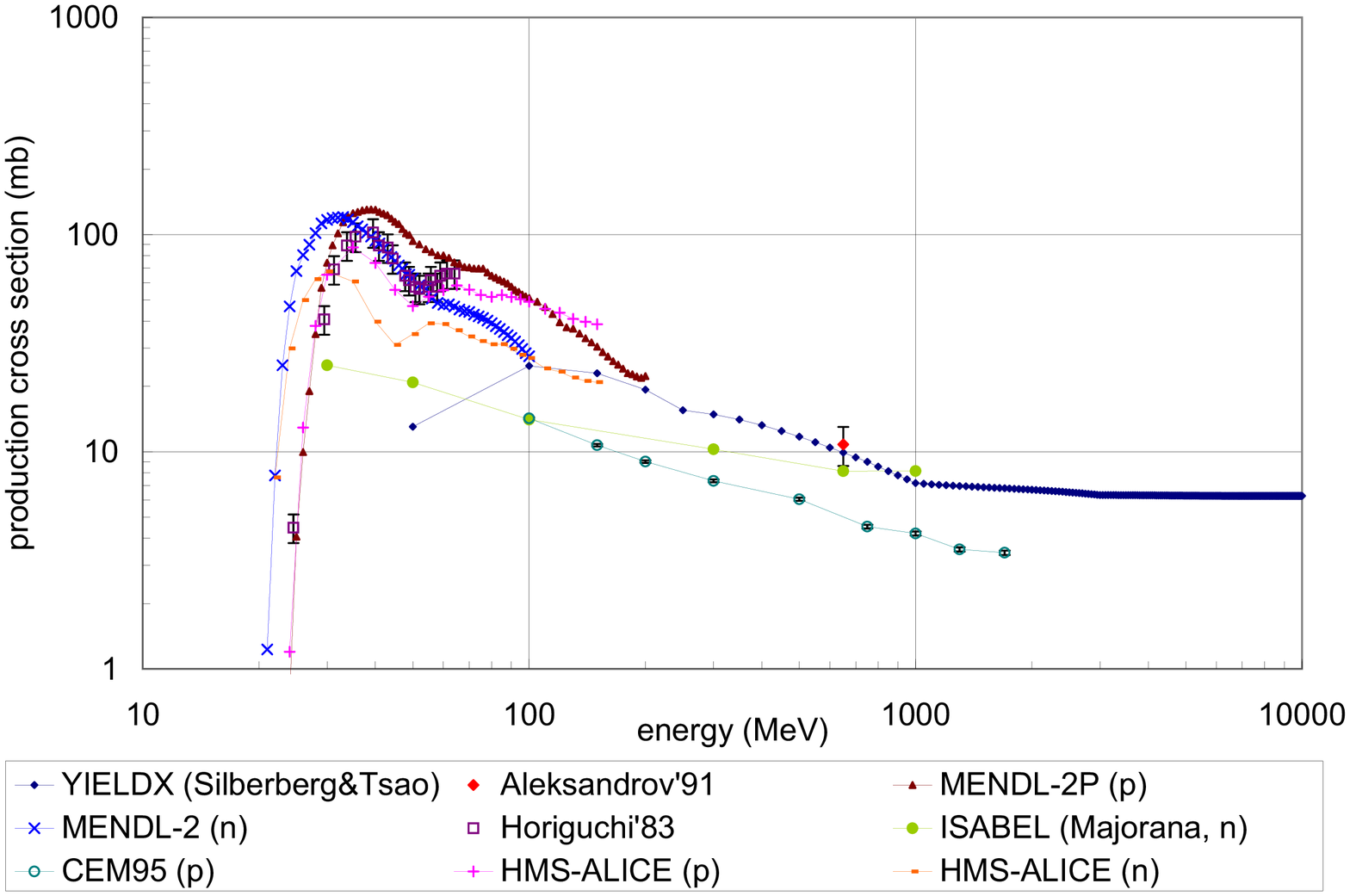}\\ 
  \includegraphics[height=.2\textheight]{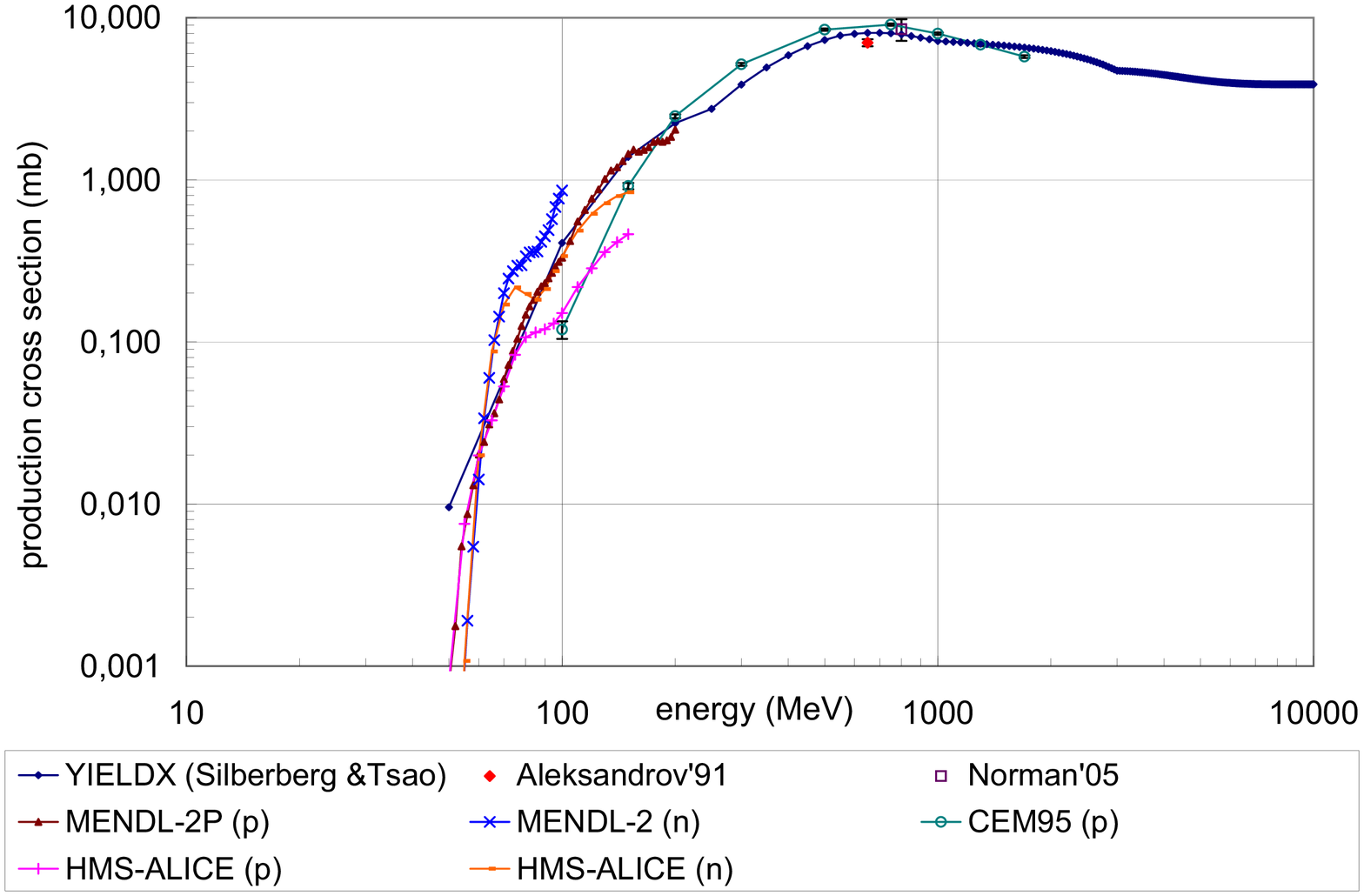}
  \caption{Comparison of excitation functions for $^{68}$Ge (top) and for $^{60}$Co (bottom) in natural germanium by nucleons from different sources:
  measurements (Horiguchi'83 \cite{horiguchi}, Aleksandrov'91 \cite{ref660M} and
Norman'05 \cite{neutrinonorman}), calculations using the YIELDX code
\cite{tsao3}, and Monte Carlos results (elaborated by the Majorana
Collaboration using ISABEL \cite{majorana}, taken from the MENDL2
libraries based on ALICE code \cite{mendl} and taken from the
library in Refs. \cite{study,la150} using CEM95 and HMS-ALICE
codes).}
  \label{ge68geco60ge}
\end{figure}

Excitation functions of $^{60}$Co and $^{68}$Ge, as well as of other
long-lived products induced in germanium, have been compiled using
available measurements and different calculations taken from
libraries or made on purpose in Ref. \cite{cosmogenicstaup2005} (see
Figure \ref{ge68geco60ge}). In an attempt to find the most reliable
selection of the excitation functions, those calculations with
minimum deviations with respect to experimental data were taken into
consideration: below 150 MeV, HMS-ALICE results for neutrons
\cite{study,la150}, and above this energy, YIELDX calculations.

Once chosen the excitation functions, production rates of $^{60}$Co
and $^{68}$Ge were calculated using the cosmic neutron spectrum from
Ziegler \cite{ziegler}, for both natural and enriched germanium
(86\% of $^{76}$Ge, 14\% of $^{74}$Ge). Results are summarized in
Tables \ref{ratesnatural} and \ref{ratesenr}, compared with some
previous estimates.

\begin{table*}
\begin{center}
\caption{Production rates (in kg$^{-1}$d$^{-1}$) in natural
germanium obtained considering HMS-ALICE and YIELDX below and
above 150 MeV respectively, together with previous estimates. Some
results from Ref. \cite{avignone} are based on Monte Carlo
calculations (MC) while others come from measurements (exp).
} \centering
\begin{tabular}{l| c r @{.} l r @{.} l r @{.} l c}
\hline
  &  HMS-ALICE+YIELDX  \cite{cosmogenicstaup2005} &   \multicolumn{2}{c}{Ref. \cite{baudis}} &  \multicolumn{2}{c}{Ref. \cite{miley}} & \multicolumn{2}{c}{Ref.
  \cite{avignone} (MC)}  & Ref. \cite{avignone} (exp) \\
  \hline
$^{68}$Ge  &  77+12=89   &  58&4  & 26&5 &  29&6 & 30$\pm$7 \\
$^{60}$Co  &  0.3+4.5=4.8  &6&6  &  4&8 & \multicolumn{2}{c}{}& \\
\hline
\end{tabular}
 \label{ratesnatural}
\end{center}
\end{table*}

\begin{table*}
\begin{center}
\caption{As Table \ref{ratesnatural}, but for enriched germanium
(86\% of $^{76}$Ge and 14\% of $^{74}$Ge).
} \centering
\begin{tabular}{l| c r @{.} l r @{.} l}
\hline
  &  HMS-ALICE+YIELDX  \cite{cosmogenicstaup2005} &  \multicolumn{2}{c}{Ref. \cite{miley}} & \multicolumn{2}{c}{Ref.
  \cite{avignone}} \\ \hline
$^{68}$Ge  & 2.8+10=13  &  1&2   &0&94 \\
$^{60}$Co  & 0.02+6.7=6.7   & 3&5 & \multicolumn{2}{c}{}\\
\hline
\end{tabular}
\label{ratesenr}
\end{center}
\end{table*}

A production rate of $\sim$5 kg$^{-1}$d$^{-1}$ can be safely
considered in natural Ge for $^{60}$Co while for $^{68}$Ge the
production rate of $\sim$90 kg$^{-1}$d$^{-1}$ found in this work
is significantly higher than in previous estimates including the
measurements in Ref. \cite{avignone} due to the large contribution
of neutrons below 100 MeV. Since there is now no available
measurement of production cross sections by neutrons for the
relevant isotopes, it is difficult to assess to what extent the
two basic assumptions of the presented calculations are valid. For
enriched detectors, production of $^{68}$Ge, although much more
reduced than the one for natural germanium,  has been found one
order of magnitude higher than previous estimates commonly used
\cite{miley,avignone}. $^{60}$Co production seems to be similar in
enriched and natural material.

\subsubsection{Counting rates in the region of interest}

A set of Monte Carlo simulations based on GEANT4 code have been
made to reproduce the response of cylindrical germanium detectors
of different masses to the background sources, including $^{60}$Co
and $^{68}$Ge isotopes induced in the germanium crystal.

The number of events registered in the 2-2.1 MeV RoI per isotope
decay can be deduced using these MC simulations. For 2(4)-kg
detectors, these numbers are 0.00016 (0.00020) c/keV/decay of
$^{60}$Co and 0.00023 (0.00029) c/keV/decay of $^{68}$Ge (in
agreement with the ones deduced in simulations made by the
Majorana \cite{majorana} and GERDA \cite{gerda} collaborations).

Using this information from GEANT4 simulations together with the
production rates previously presented, the counting rates in the
region of interest due to the cosmogenic contaminations can be
derived for certain exposure and cooling times of the material. To
properly compare results, the same times typically used in the
GERDA project \cite{gerdataup} will be considered: 30 days of
exposure to cosmics rays for $^{60}$Co production and 180 days of
exposure plus 180 days of cooling for $^{68}$Ge. Table
\ref{countingrates} presents the obtained counting rates in these
conditions and using production rates of 5 kg$^{-1}$d$^{-1}$ for
$^{60}$Co and 1 kg$^{-1}$d$^{-1}$ for $^{68}$Ge; it is also shown
the estimate when considering a $^{68}$Ge production rate of 10
kg$^{-1}$d$^{-1}$ and 2 years of cooling. As it can be seen, for
this high $^{68}$Ge production rate it would be necessary to wait
more than two years (instead of 3 months) to achieve counting
rates of the same order than when considering the low production
rate. Rates presented in Table \ref{countingrates} correspond to
conventional germanium detectors, that is, without taking into
account neither segmentation nor PSA techniques.

\begin{table*}
\begin{center}
\caption{Estimates of counting rates R (in units of $10^{-3}$ c
keV$^{-1}$ kg$^{-1}$y$^{-1}$) in the 2-2.1 MeV RoI from cosmogenic
contaminations in 2-kg and 4-kg germanium detectors. Uncertainties
in these estimates are discussed in the text.} \centering
\begin{tabular}{l|ccccccc}
\hline
& production rate && exposure && cooling & R & R \\
& (kg$^{-1}$d$^{-1}$) && (d) && (d) & 2 kg & 4 kg\\
\hline
$^{60}$Co & 5 && 30 && 0 & 2.9 & 3.7 \\
$^{68}$Ge & 1 && 180 && 180 & 12 & 16 \\
$^{68}$Ge & 10 && 180 && 730 & 31 & 39\\
\hline
\end{tabular}
\label{countingrates}
\end{center}
\end{table*}

\subsection{2614.5 keV $^{208}$Tl emissions}

An evaluation of the expected background coming from the 2614.5
keV line from $^{208}$Tl has been made, considering both the
environmental gamma background in the laboratory and the
$^{232}$Th intrinsic radioimpurities in the lead shielding
expected to be surrounding detectors. The response to 2614.5 keV
photons for the counting rates in the 2-2.1 MeV RoI in 2-kg and
4-kg germanium detectors has been estimated by Monte Carlo
simulation. A flux of $\sim$0.1 cm$^{-2}$ s$^{-1}$ for
environmental 2614.5 keV photons has been assumed, according to
recent measurements in the new Canfranc Underground Laboratory
\cite{gammacanfranc}, and an activity of 1 $\mu$Bq/kg of
$^{232}$Th in lead has been considered just as a reference value.
In these estimates, a spherical cavity with radius R=30 cm for
placing detectors inside has been supposed, and two different
shielding configurations with 30 and 40 cm of lead surrounding the
cavity have been taken into account. Table \ref{countingratesTl}
summarizes the obtained results.

\begin{table*}
\begin{center}
\caption{Estimates of counting rates (in units of $10^{-3}$ c
keV$^{-1}$ kg$^{-1}$y$^{-1}$) in the 2-2.1 MeV RoI from 2614.5 keV
$^{208}$Tl emissions in 2-kg and 4-kg germanium detectors (see
text). Two shielding configurations with 30 and 40 cm of lead have
been taken into consideration. Uncertainties in these estimates
are discussed in the text.} \centering
\begin{tabular}{l|cc}
\hline
& 2 kg & 4 kg \\

\hline
external gamma, 30 cm Pb & 40 & 32 \\
external gamma, 40 cm Pb & 0.38 & 0.30 \\
intrinsic radioimpurities in Pb& 2.8 & 2.2\\
\hline
\end{tabular}
\label{countingratesTl}
\end{center}
\end{table*}

For the intrinsic radioimpurities in the lead shielding, it has
been checked that around 90\% of the events come from the most
internal 5 cm of shield, and in fact, virtually the same counting
rates have been found when considering 30 or 40 cm of lead
(despite the significant difference in mass and consequently in
$^{232}$Th activity between the two configurations). For the
external gamma background, the additional 10 cm when assuming a
40-cm-thick lead layer reduce the counting rates around two orders
of magnitude. Comparing the background levels expected in 2 and
4-kg detectors in Tables \ref{countingrates} and
\ref{countingratesTl}, the largest crystals have a lower
background level ($\sim$20\%) from $^{208}$Tl but higher counting
rates ($\sim$30\%) from internal cosmogenic impurities. This fact
will be discussed at Sec. \ref{granularity}.

Uncertainties in the estimates of raw backgrounds presented in
Tables \ref{countingrates} and \ref{countingratesTl} come in
principle from the simulations of the response of detectors to the
different background sources as well as from the inputs considered
for production rates of cosmogenic isotopes and levels of
$^{208}$Tl photons. Counting rates are directly proportional to
these inputs, which are thought to be the main source of error
since GEANT4 can reproduce electromagnetic processes with a few
per cent error and statistical errors in simulations have always
been kept below 2\% as stated before. Production rate of $^{60}$Co
has an uncertainty of $\sim$50\% (when considering different
available estimates) and for $^{68}$Ge it could be up to one order
of magnitude. Environmental gamma fluxes depend on the particular
underground location. For $^{232}$Th impurities in very pure lead
just upper bounds have been derived (see Refs.
\cite{heusserlrt,laubenstein}) and therefore the assumed value
must be taken just as a reference.

\section{Rejection of background events} \label{rejection}


The main goal in the study of the different ways to reduce the
background was to try to analyze the correlation between the
experiment design and background level in the region of interest.
Assuming a hypothetical final set up of an experiment with a total
mass of germanium around some tens of kilograms, the aim was to
determine what is the optimal mass distribution and features of
detectors to have a background level as low as possible. Three
topics that can determine the best configuration, as we pointed
out before, are: granularity of the experiment detectors,
segmentation of the crystals and analysis of the obtained pulses.

\subsection{Granularity} \label{granularity}


The first step to determine the best set up of the experiment
consists in analyzing what could be the optimal mass of the
detectors that build the whole experiment to have a background
level as low as possible, if a fixed total mass of germanium is
assumed. With this purpose, cylindrical crystals with the same
value for diameter and height and masses between 0.1 and 4 kg have
been simulated.

\begin{figure}
\centering
  \includegraphics[height=.2\textheight]{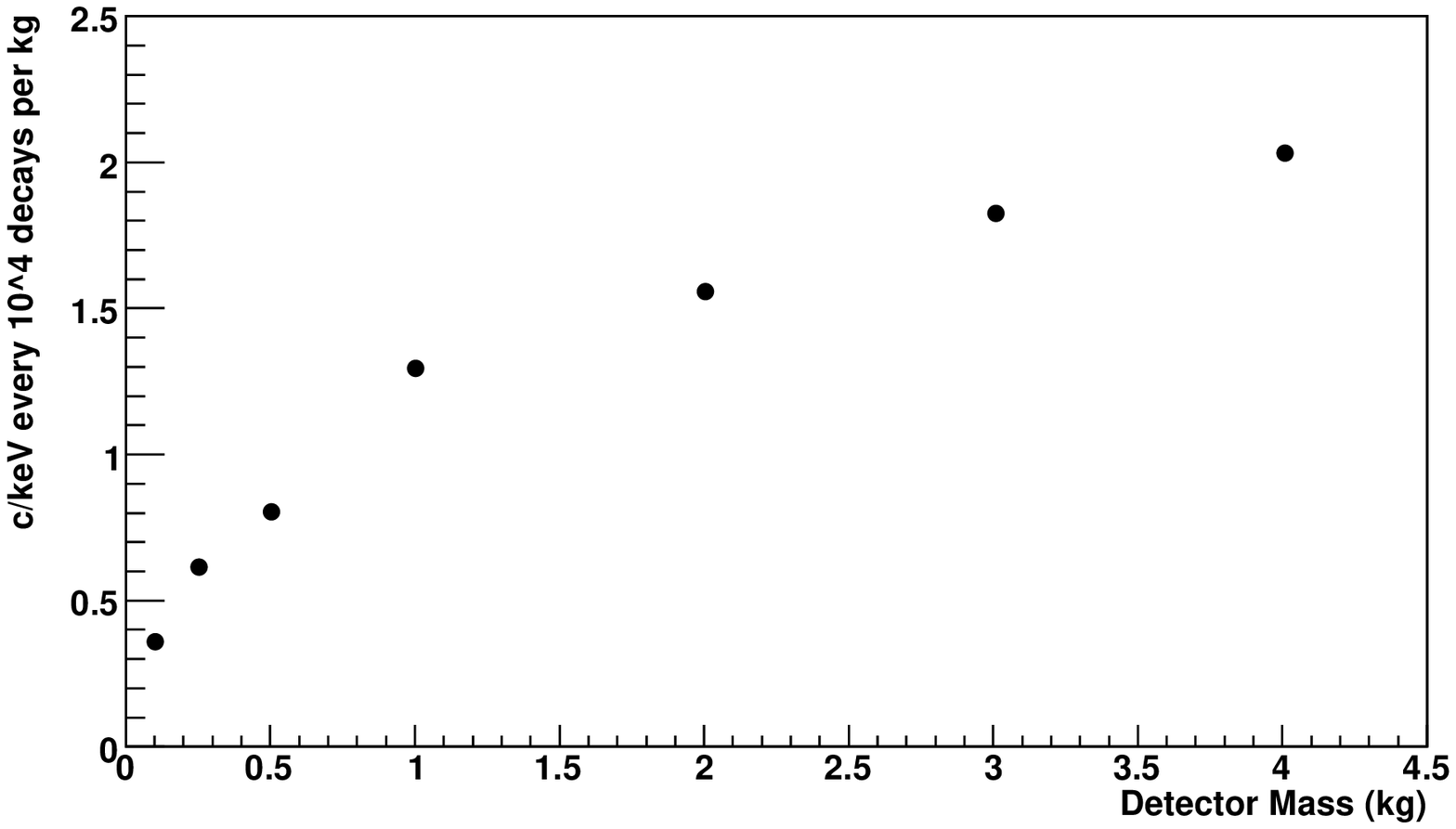} \\a)\\
  \includegraphics[height=.2\textheight]{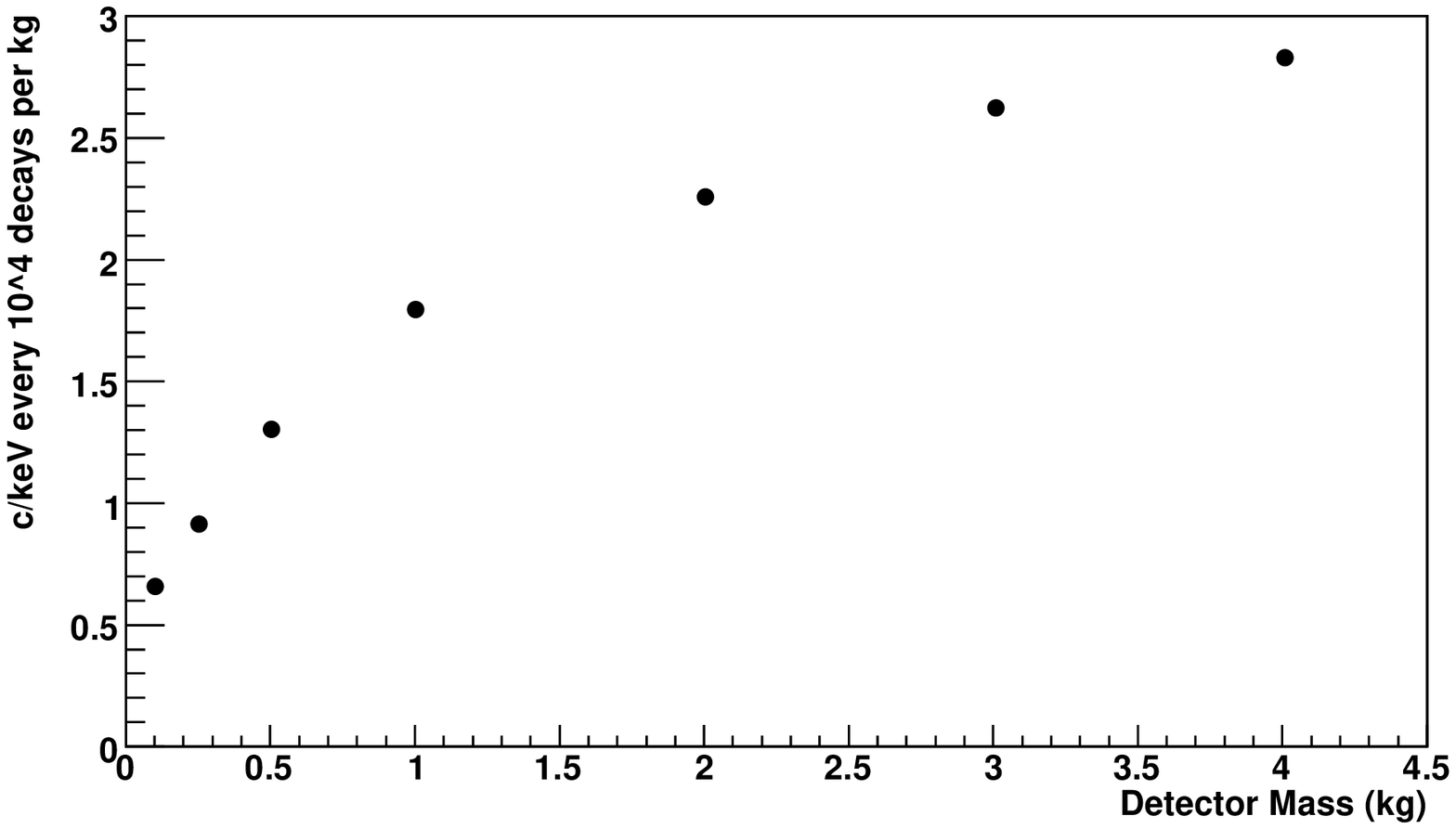}\\b)\\
  \includegraphics[height=.2\textheight]{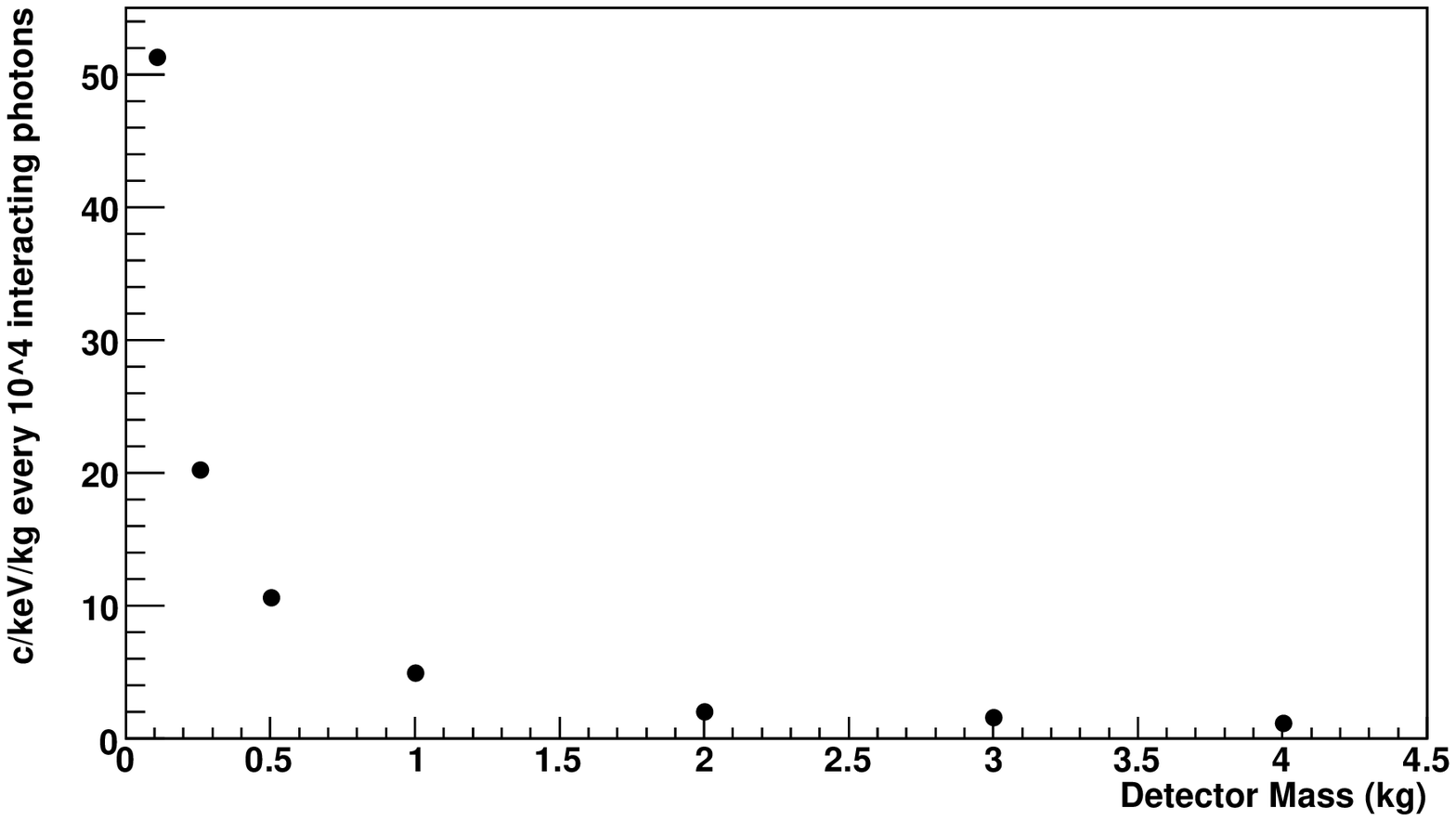}\\c)
  \caption{Background level in the 2-2.1 MeV RoI depending on the component detector mass. For internal
  contaminations,
   $^{60}$Co (a) and $^{68}$Ge (b), it is represented in counts per keV every 10$^{4}$ decays per kg. For external photons of 2614.5 keV
   (c), it
   is represented in counts per keV per kg every 10$^{4}$ interacting photons.}
  \label{Granularity}
\end{figure}

If only the events with an energy deposit between 2 and 2.1 MeV
are taken into account, we can see how for internal contaminations
and a given specific activity, the higher mass detectors register
a higher background level (see Figs.\ref{Granularity} a,b). 4-kg
detectors register 28\% (26\%) more events for $^{60}$Co
($^{68}$Ge) than 2-kg ones. This is not the case for external
contaminations, because for a given activity, the heavier
detectors have a lower background (see Fig.\ref{Granularity} c).
4-kg detectors register 54\% less events coming from 2614.5 keV
photons than 2-kg detectors. These values confirm that the optimal
configuration of an experiment depends on what kind of background
we want to reduce more, that coming from internal contaminations
or from external ones, taking also into account that the
dependency between the mass of the detectors and the background
events registered is stronger for the external contamination.
These data together with conclusions obtained from the study of
the segmentation of the crystal and pulse analysis, as explained
later, will determine the best configuration.

\subsection{Segmentation}

\begin{figure}
\centering
  \includegraphics[height=.25\textheight]{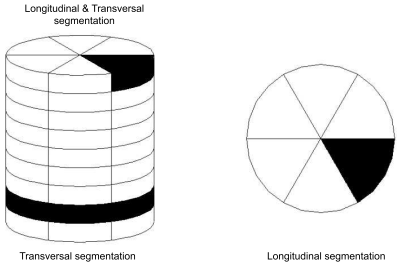}
  \caption{Segmentation scheme for a germanium crystal with 9 transversal slices and 6 longitudinal sectors to obtain 54 segments.}
  \label{EsqS_pap}
\end{figure}

Regardless of the mass of the used detectors, other way to reduce
the background level is by segmentation of the crystals and
further application of anticoincidence techniques between
segments. The aim was to quantify the maximum background reduction
that could be obtained from segmentation of the crystals. For the
cylindrical detectors studied, two different ways to divide it
were considered: segmentation in planes and segmentation in
sectors (longitudinal and transversal segmentation respectively)
(see Fig. \ref{EsqS_pap}).

\begin{figure}
\centering
 \includegraphics[height=.2\textheight]{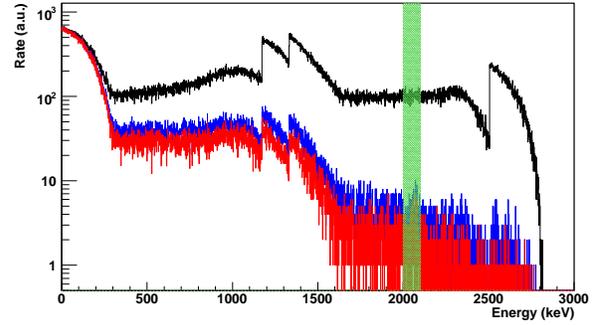} \\a)\\
 \includegraphics[height=.2\textheight]{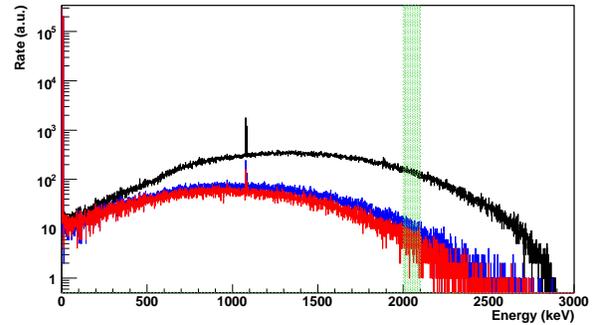}\\b)\\
 \includegraphics[height=.2\textheight]{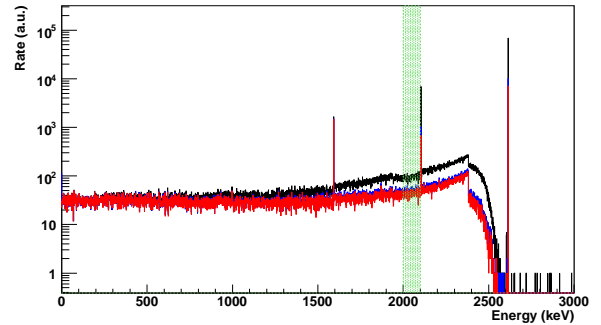}\\c)
  \caption{Comparison between the spectra obtained for the different contaminations studied: $^{60}$Co (a), $^{68}$Ge (b) and 2614.5 keV photons (c), for a
  4-kg detector without segmentation (black line), with 11 transversal segments (blue line) and with 66 segments, 11 transversal by 6 longitudinal (red line).}
  \label{Segm_pap}
\end{figure}

Table \ref{reductionsegmentation} shows how the reduction of
counts registered between 2 and 2.1 MeV in detectors of 2 and 4-kg
is bigger for higher number of segments, combining longitudinal
and transversal segmentation. For a 4-kg detector with the highest
number of segments studied (66 segments distributed in 11
transversal slices and 6 angular sectors), approximately only 2
out of 100 events will not be rejected by anticoincidence
techniques for $^{60}$Co contamination. For $^{68}$Ge the ratio is
around 5 out of 100 events and for 2614.5 keV external gammas,
less than a half of total events. This reduction can be observed
qualitatively in the spectra obtained from simulations for the
different background sources studied (see Fig. \ref{Segm_pap}).

\begin{table*}
\begin{center}
\caption{Percentage of rejected events by crystal segmentation and
anticoincidence techniques for registered events between 2 and 2.1
MeV. Different segmentation schemes for 2 and 4-kg detectors have
been considered.} \centering
\begin{tabular}{l|cccc|cccc}
\hline
 &\multicolumn{4}{c|}{2kg}&\multicolumn{4}{c}{4kg}\\
 &  &  & 7 planes & 9 planes &  &  & 9 planes & 11 planes \\
&&& $\&$ & $\&$ &&& $\&$ & $\&$ \\
& 7 planes& 9 planes& 6 sectors & 6sectors &9 planes&11 planes& 6 sectors & 6 sectors \\
 \hline
$^{60}$Co   & 92.4 & 95.6 & 97.0 & 98.2 & 94.9 & 96.7 & 97.6 & 98.4 \\
$^{68}$Ge   & 86.1 & 90.7 & 93.2 & 95.1 & 89.8 & 92.7 & 94.2 & 95.7 \\
external 2614.5 keV gammas& 40.6 & 44.4 & 48.6 & 51.0 & 45.4 & 49.4 & 52.4 & 55.1 \\
 \hline
\end{tabular}
\label{reductionsegmentation}
\end{center}
\end{table*}

It is clear that for higher number of segments in the crystal, we
are able to reject more background events. But the segmentation of
a crystal is limited for some reasons. First of them is the
difficulty associated to the reduction of the width of the
transversal segments. It is also necessary to make a segmentation
to reject a high number of background events without losing too
much efficiency in the double beta decay events detection. The
different segmentation configurations studied have transversal
segments with a width of 1 cm approximately. This size is
reachable using actual segmentation techniques and provides a good
efficiency for the detection of double beta decay events, as it
will be discussed later in Sec. \ref{Efficiency to signal}.

Considering the biggest segmentation of all studied and applying
the reduction factors obtained and shown in Table
\ref{reductionsegmentation} to the raw rates for different
background sources estimated in Sec. \ref{raw}, the resulting
rates can be derived and are presented in the first and third
columns of Table \ref{reducedcountingrates}.


\begin{table*}
\begin{center}
\caption{Estimates of counting rates (c keV$^{-1}$
kg$^{-1}$y$^{-1}$) in the 2-2.1 MeV RoI from cosmogenic
contaminations and 2614.5 keV gamma emissions from $^{208}$Tl in
2-kg and 4-kg germanium detectors, assuming maximum reduction
factors deduced for segmented crystals using anticoincidence
techniques and for segmented crystals using PSA. Same conditions
for cosmogenic production and 2614.5 keV emissions than in Tables
\ref{countingrates} and \ref{countingratesTl} are taken into
account.} \centering
\begin{tabular}{l|cc|cc}
\hline
&\multicolumn{2}{|c|}{2kg}& \multicolumn{2}{c}{4kg}   \\
 & segmentation & segmented PSA   & segmentation & segmented PSA\\
\hline
$^{60}$Co   &  5.2$\times 10^{-5}$ & 8.7$\times 10^{-6}$ &  6.0$\times 10^{-5}$ & 3.7$\times 10^{-6}$ \\
$^{68}$Ge   &   6.1$\times 10^{-4}$ & 1.4$\times 10^{-4}$ &   6.8$\times 10^{-4}$ & 1.3$\times 10^{-4}$ \\
$^{68}$Ge  &  1.5$\times 10^{-3}$ & 3.4$\times 10^{-4}$ &   1.7$\times 10^{-3}$ & 3.1$\times 10^{-4}$ \\
external 2614.5 keV, 30 cm Pb& 1.9$\times 10^{-2}$ & 1.8$\times 10^{-2}$ & 1.4$\times 10^{-2}$ & 1.3$\times 10^{-2}$ \\

external 2614.5 keV, 40 cm Pb &  1.8$\times 10^{-4}$ & 1.7$\times
10^{-4}$ & 1.4$\times
10^{-4}$ & 1.2$\times 10^{-4}$ \\
intrinsic 2614.5 keV in lead&  1.4$\times 10^{-3}$ & 1.2$\times 10^{-3}$ &  1.0$\times 10^{-3}$ & 8.9$\times 10^{-4}$ \\
 \hline
 Best Total & 2.2$\times 10^{-3}$ & 1.6$\times10^{-3}$& 1.9$\times 10^{-3}$ & 1.1$\times 10^{-3}$ \\
 \hline
\end{tabular}
\label{reducedcountingrates}
\end{center}
\end{table*}

\subsection{PSA}
\label{PSA_section}


Besides the improvements made in the detector, like the
segmentation of the crystal, a reduction of the background level
from the analysis of the pulses registered can also be obtained.
For this purpose, as pointed out before, it is necessary to
distinguish between background and real double beta decay events
in order to reject the first ones. Application of PSA in segmented
crystals allows to increase the spatial resolution of conventional
germanium detectors to obtain the position of energy deposits with
a very good accuracy in all the dimensions \cite{genetic} or at
least a correct identification of the number of interaction points
\cite{crespi}.

\begin{figure}
\centering
 \includegraphics[height=.2\textheight]{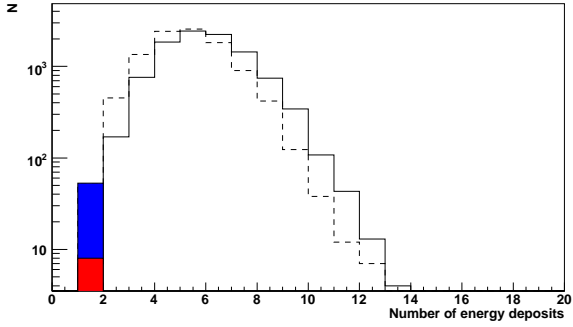} \\a)\\
 \includegraphics[height=.2\textheight]{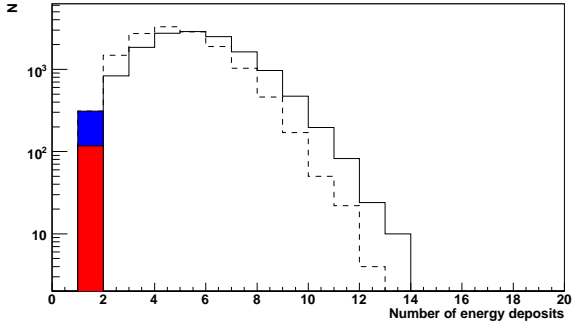}\\b)\\
 \includegraphics[height=.2\textheight]{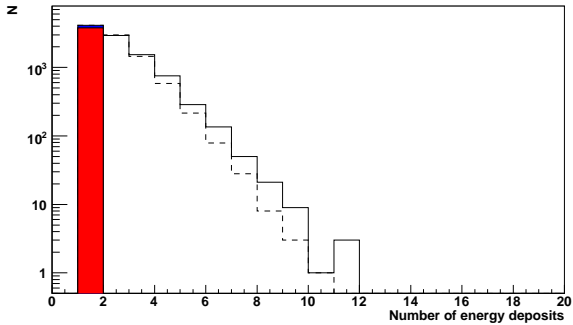}\\c)
  \caption{Distribution of the number of energy deposits per event in the 2-2.1 MeV RoI for all the background contributions studied:
  $^{60}$Co (a), $^{68}$Ge (b) and 2614.5 keV photons (c), for 4-kg detector and considering a spatial resolution of 3 (solid line) and 5 mm (dashed line). Monosite events are singled out in red (blue) for 3 (5) mm resolution.}
  \label{3-5mm_Pap}
\end{figure}

The data obtained in the simulations can be reanalyzed for a given
spatial resolution, grouping all the partial energy deposits with
a separation lower than the resolution and considering these
groups like indivisible energy deposits. Then, it is possible to
determine how many of these indivisible deposits each background
event has. In figure \ref{3-5mm_Pap}, distributions of the number
of energy deposits per event for different background sources and
4-kg detectors are showed, assuming 2 different values for the
spatial resolution, 3 and 5 mm, which seem to be at reach today
according to the work developed in Ref. \cite{genetic}\footnote{A
genetic algorithm for the decomposition of multiple hit events is
presented in Ref. \cite{genetic}, considering the features of a
cylindrical closed-end germanium detector with a mass of almost 2
kg and 6 angular sectors and 4 transversal slices. However, it is
reported that the approach has no limitation concerning the
geometry of the crystal, the number and layout of the segments or
the number of interactions.}; it is worth noting that these
distributions have been obtained without making any particular
definition of the segmentation scheme. The worst value of the
spatial resolution can give an idea of the loss of rejection
efficiency depending on the spatial resolution finally achieved.
In Table \ref{reductionPSA}, rejection factors after eliminating
"multisite" events, between 2 and 2.1 MeV, are showed in function
of the 2 values considered for the spatial resolution and for all
background sources studied and 2 and 4-kg detectors. For the
heaviest crystal and assuming a 3 mm spatial resolution, it is
possible to reject 99.9\% of background events coming from
$^{60}$Co, 99.2\% from $^{68}$Ge and 60.3\% from 2614.5 keV
photons; the values are 99.5\%, 97.8\% and 56.5\% respectively if
a resolution of 5 mm is considered.

\begin{table*}
\begin{center}
\caption{Percentage of rejected events in the 2-2.1 MeV RoI by PSA
techniques considering a spatial resolution of 3 and 5 mm for 2
and 4-kg detectors.} \centering
\begin{tabular}{l|cc|cc}
\hline
&\multicolumn{2}{|c|}{2kg}& \multicolumn{2}{c}{4kg}  \\
&3 mm & 5 mm & 3 mm & 5 mm \\
&resolution& resolution&resolution& resolution \\
 \hline
$^{60}$Co   & 99.7 & 99.0 & 99.9 & 99.5 \\
$^{68}$Ge   & 98.9 & 97.0 & 99.2 & 97.8 \\
external 2614.5 keV gammas& 55.6 & 51.2 & 60.3 & 56.5 \\
 \hline
\end{tabular}
\label{reductionPSA}
\end{center}
\end{table*}

\begin{figure}
\centering
  \includegraphics[height=.2\textheight]{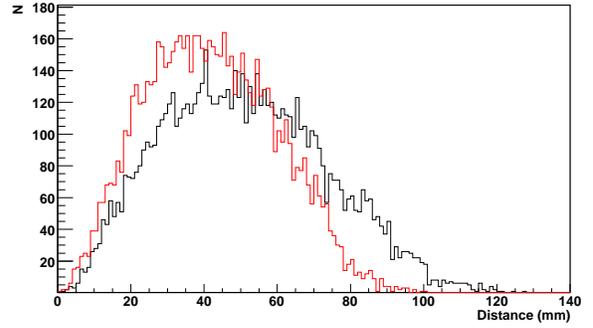} \\a)\\
  \includegraphics[height=.2\textheight]{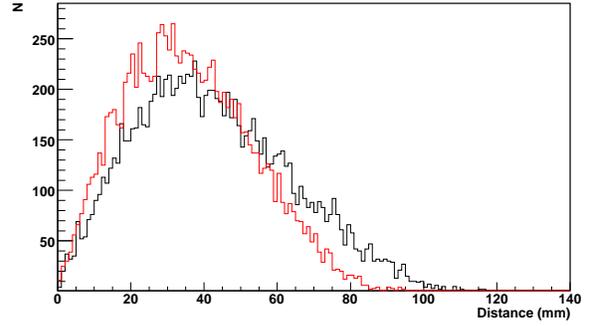}\\b)\\
  \includegraphics[height=.2\textheight]{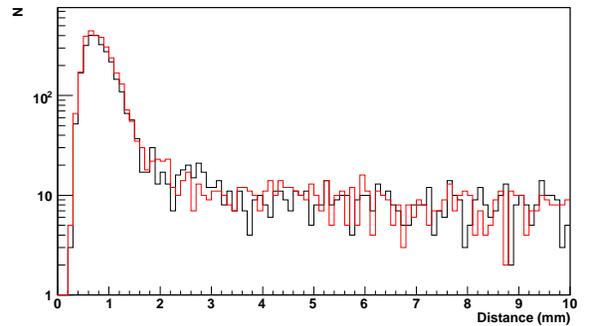}\\c)
  \caption{Distribution of the maximum interdistance between energy deposits of the same event in the 2-2.1 MeV RoI for $^{60}$Co (a)and $^{68}$Ge (b)
  internal contaminations and for 2614.5 keV external photons (c), for  2-kg (red line) and 4-kg (black line) detectors.}
  \label{Dmax_Pap}
\end{figure}

Spatial resolution of the detectors allows to reject more or less
background events depending on how the energy of these events is
distributed in the crystal. One way to predict what could be the
maximum reduction factor consists in determining the maximum
distance between all the energy deposits of the same event (that
we call maximum interdistance $D_{max}$, see Eq. \ref{DmaxEq}) and
studying how it is distributed.
\begin{equation}
D_{max}=max[\sqrt{(x_{i}-x_{j})^{2}+(y_{i}-y_{j})^{2}+(z_{i}-z_{j})^{2}}]
      \label{DmaxEq}
\end{equation}
This maximum interdistance depends on different factors like the
mass of the detector, the scheme of the decay of the isotope that
produces the background event or the origin of the event, because
it is different if the contamination is located inside the crystal
or if comes from outside of the experimental setup. In Fig.
\ref{Dmax_Pap}, the distribution of these maximum interdistances
for 2 and 4-kg detectors and all the background sources studied
are presented, showing how all the factors mentioned previously
have influence on these distributions. From these plots in Fig.
\ref{Dmax_Pap}, rejection factors for any considered experimental
spatial resolution can be deduced knowing that all events with a
maximum interdistance lower than the spatial resolution will be
labelled as "monosite" like double beta events.

Second and fourth columns in Table \ref{reducedcountingrates} show
the expected counting rates in the 2-2.1 MeV RoI for background
contributions in 2-kg and 4-kg germanium detectors, applying the
reduction factors obtained when considering a 3D spatial
resolution of 3 mm from PSA on the raw backgrounds estimates in
Sec. \ref{raw}. Total background levels have been calculated (see
last row in Table \ref{reducedcountingrates}) including cosmogenic
$^{60}$Co and $^{68}$Ge in the detector as well as external and
lead 2614.5 keV emissions and using the most favorable conditions,
that is, adding values in first, second, fifth and sixth rows.

It must be noted that no technical limitation in the process of
rejecting events thanks to coincidences between segments or PSA
with a certain spatial resolution has been taken into account.
Consequently, the rejection factors summarized in Tables
\ref{reductionsegmentation} and \ref{reductionPSA} must be
considered as the best ones achievable and give the limit of the
power of these rejection techniques. Uncertainties in the total
background levels presented in Table \ref{reducedcountingrates}
must be dominated by the uncertainties in the raw backgrounds,
discussed in Sec. \ref{raw}, since statistical errors in the
simulations are much smaller.

\section{Efficiency to neutrinoless DBD signal and sensitivity}
\label{sensitivity}

The sensitivity of a neutrinoless DBD experiment is often evaluated
using the detector factor-of-merit F$_{D}$ defined as:
\begin{equation}
F_{D}=4.17\times10^{26} \frac{f}{W}\sqrt{\frac{MT}{b\Delta
E}}\epsilon \label{detectorfom}
\end{equation}
with f the isotopic abundance of the DBD emitter, W the atomic
weight of the source material, MT the exposure of the experiment,
b the background level (typically expressed in counts per keV, per
kg and per y), $\Delta E$ the energy window where the signal is
expected to appear, dependant on the energy resolution of the
detector, and $\epsilon$ the detection efficiency. F$_{D}$ is
interpreted as the lifetime corresponding to the minimum
detectable number of events over a background at 1$\sigma$
confidence level.

The neutrino effective masses which can be explored by an
experiment with a detector factor-of-merit F$_{D}$ can be
determined as:
\begin{equation}
<m_{\nu}>^{2}= \frac{m_{e}^{2}}{F_{D} F_{N}}\label{mef}
\end{equation}
with m$_{e}$ the electron mass and F$_{N}$ the nuclear
factor-of-merit, defined as F$_{N}$=G$^{0\nu}|$M$^{0\nu}|^{2}$,
being G$^{0\nu}$ a kinematical factor and M$^{0\nu}$ the nuclear
matrix element qualifying the likeliness of the transition.

The background level $b$ achievable in germanium experiments has
been analyzed in previous sections under different background
reduction schemes; but the application of anticoincidence
rejection or PSA techniques unfortunately affects also the
efficiency for the detection of the neutrinoless DBD signal.
Therefore, a study of this efficiency has been carried out and is
presented here.

\subsection{Efficiency to signal} \label{Efficiency to signal}

In order to evaluate the dependency between background reduction
techniques and loss of efficiency, it is necessary to simulate
neutrinoless DBD ($0\nu\beta\beta$) events inside the detector to
apply them the same treatment that a background event. This
procedure allows to estimate the percentage of $0\nu\beta\beta$
events rejected losing detection efficiency. Signal events can be
missed either because of the escape of the Bremsstrahlung
radiation produced by electrons or because $0\nu\beta\beta$ events
are mistaken as background due to the spatial distribution of
their energy deposits.

To define $0\nu\beta\beta$ events in the simulations, two
electrons with sum energy of 2040 keV (Q value for $^{76}$Ge DBD)
have been considered, neglecting in first approach the angular
correlation between electrons. In order to check if this angular
correlation could have substantial influence to detect a
$0\nu\beta\beta$ event, a first study was made simulating two
electrons of 1020 keV each emitted randomly, with the same and
with opposite directions. Decays are distributed homogenously in
the crystals, for 2 and 4-kg detectors. Tables \ref{2kg-1020} and
\ref{4kg-1020} show that the differences between the detection
efficiency factors in the cases described before are less than 1\%
for the different angular correlations considered, validating the
approximation taken.

Another point in the simulation of $0\nu\beta\beta$ events is the
energy distribution of the two electrons. Four different
configurations were studied: two electrons of 1020 keV each (half
of the total energy), two electrons of 1500 and 540 keV, two
electrons of 1734 and 306 keV (75\% and 25\% of the full energy
respectively) and one electron of 2040 keV. It is important to
point that the last case is not a real case but could be useful to
obtain limit values in the detection efficiency because
Bremsstrahlung probability is higher for more energetic electrons.
For all these energy configurations, the electrons were emitted in
random directions using the same simulation package that for the
background events. Figure \ref{DBD_1020} shows an example of the
spectrum registered in the detector after the simulation of
$0\nu\beta\beta$ events. In this case in particular, two electrons
of 1020 keV each were emitted in a 4-kg detector. Three different
regions can be identified in this spectrum:

\begin{figure}
\centering
  \includegraphics[height=.2\textheight]{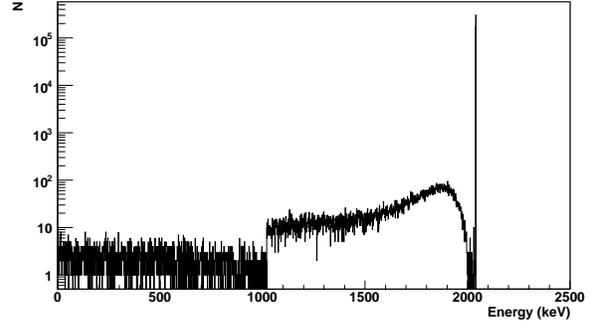}
  \caption{Energy spectrum registered for simulated $0\nu\beta\beta$ events composed by 2 electrons of 1020 keV each emitted in random directions in a 4-kg detector.}
  \label{DBD_1020}
\end{figure}

\begin{itemize}
\item Up to 1020 keV, energy is lost from both electrons, either
by escape of Bremsstrahlung radiation and/or escape of the
electron itself (if the decay is close enough to the detector
surface). The probability that this happens for both electrons is
quite low and for this reason a step appears at 1020 keV.

\item From 1020 keV up to the peak, events where one electron
deposits all the energy and the other one suffers Bremsstrahlung
losses or escapes from the detector, are registered. In a smaller
percentage, events where both electrons do not deposit the full
energy can appear here too.

\item In the peak, electrons have deposited the full energy. This
is the region where the $0\nu\beta\beta$ signal is expected. As
explained later, some events in the peak can be labelled as
"multisite" events and therefore rejected as background, when
electrons produce Bremsstrahlung emission which is finally
absorbed in the crystal.
\end{itemize}

Using the simulation of $0\nu\beta\beta$ events, the relationship
between the detection efficiency and the detector mass, directly
related with the granularity of a future experiment, can be
analyzed. Tables \ref{2kg-1020} and \ref{4kg-1020} show, for two
1020 keV electrons, that the difference between the detection
efficiency for 2 and 4-kg detectors is less than 1\% for all the
background rejection configurations, but always better for the
heaviest detector. For this reason, the studies of the different
energy configurations were made for a 4-kg crystal.

\begin{figure}
\centering
  \includegraphics[height=.2\textheight]{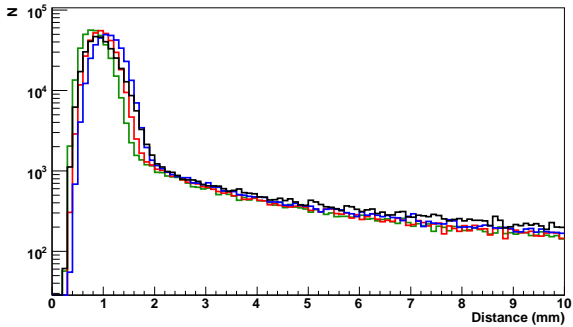}
  \caption{Distribution of the maximum interdistance between energy deposits of the same $0\nu\beta\beta$ event at the peak in a 4-kg detector
   for 2 electrons of 1020 keV each (green), 1500 + 540 keV electrons (red), 1734 + 306 keV electrons (blue) and one 2040 keV electron (black). }
  \label{Dmax_DBD}
\end{figure}

Table \ref{DBD-4kg} summarizes the detection efficiencies obtained
in a 4-kg detector for the different energy configurations in
$0\nu\beta\beta$ events. Two main conclusions can be obtained from
this table. One is that PSA techniques offer better detection
efficiency than background rejection based just on crystal
segmentation. The explanation is that the anticoincidence between
segments rejects all the $0\nu\beta\beta$ events with energy
deposits in two or more segments of the crystal, while PSA rejects
events with two separated energy deposits, typically due to
Bremsstrahlung but, in principle, could allow events close to the
borders. A useful information to understand this is the maximum
distance between all the energy deposits obtained in the
simulation for $0\nu\beta\beta$ events (the maximum interdistance
defined in Section \ref{PSA_section} and calculated using Eq.
\ref{DmaxEq}). Figure \ref{Dmax_DBD} shows the distribution of the
interdistances for signal events and for all the energy
configurations studied in a 4-kg detector. In all the cases, most
of the events have a maximum interdistance below 3 mm, ensuring
that they will be considered as "monosite" events by PSA and can
be separated safely from background ones.

\begin{figure}
\centering
  \includegraphics[height=.2\textheight]{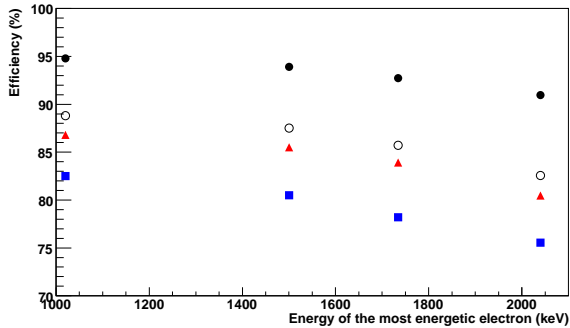}
  \caption{Detection efficiency versus the energy of the most energetic electron of a $0\nu\beta\beta$ event in a 4-kg detector
   considering a full crystal detector (black points), anticoincidences in 11 x 6 segmented crystals (blue squares), and PSA with 3 mm (red triangles) and 5 mm (white circles)
   spatial resolution.}
  \label{Eff_lines}
\end{figure}

The other important point from Table \ref{DBD-4kg} is the
dependency between the efficiency and the energy of the most
energetic electron of the $0\nu\beta\beta$ event. The more equal
the electron energies, the higher the detection efficiency. This
relationship is logical due to the Bremsstrahlung probability,
that grows proportional to the energy of the electron. For this
reason, the non real case of a 2040 keV electron can be useful to
estimate the lower limit of the detection efficiency. The obtained
dependencies of the detection efficiency on electron energy can be
fitted to a grade two polynomial, for the different background
rejection configurations considered (as shown in
Fig.\ref{Eff_lines}); convoluting this polynomials with the single
electron spectrum of $0\nu\beta\beta$ decays in $^{76}$Ge, an
overall value for the efficiency to signal can be estimated. Last
row in Table \ref{DBD-4kg} presents these results.

\begin{table*}
\begin{center}
\caption{Detection efficiency (\%) for $0\nu\beta\beta$ events
composed by two electrons of 1020 keV each emitted in random,
opposite and same directions. Different background rejection
configurations are considered, for 2-kg detectors.} \centering
\begin{tabular}{l|c|cccc|cc}
\hline
&&\multicolumn{4}{|c|}{segmentation}& \multicolumn{2}{c}{PSA}  \\
&&&&7 planes&9 planes&&\\
&&&&$\&$&$\&$&&\\
&full crystal&7 planes&9 planes&6 sectors&6 sectors&3 mm&5 mm\\
\hline
random direction&93.6&86.1&84.3&82.9&81.3&86.4&88.3\\
same direction&93.8&86.7&84.9&83.8&82.1&86.6&88.5\\
opposite direction&93.6&85.7&83.8&82.3&80.5&86.4&88.2\\
\hline
\end{tabular}
\label{2kg-1020}
\end{center}
\end{table*}

\begin{table*}
\begin{center}
\caption{As table \ref{2kg-1020}, but for 4-kg detectors.}
\centering
\begin{tabular}{l|c|cccc|cc}
\hline
&&\multicolumn{4}{|c|}{segmentation}& \multicolumn{2}{c}{PSA} \\
&&&&9 planes&11 planes&&\\
&&&&$\&$&$\&$&&\\
&full crystal&9 planes&11 planes&6 sectors&6 sectors&3 mm&5 mm\\
\hline
random direction&94.7&86.4&84.8&83.7&82.4&86.8&88.7\\
same direction&94.8&86.9&85.5&84.5&83.2&87.0&88.8\\
opposite direction&94.6&85.9&84.3&83.1&81.7&86.9&88.7\\
\hline
\end{tabular}
\label{4kg-1020}
\end{center}
\end{table*}

\begin{table*}
\begin{center}
\caption{Detection efficiency (\%) for $0\nu\beta\beta$ events
composed by two electrons emitted in random directions with
different energy schemes. Different background rejection
configurations are considered for 4-kg detectors. Last row shows
an overall efficiency taking into account the single electron
$0\nu\beta\beta$ spectrum.} \centering
\begin{tabular}{l|c|cccc|cc}
\hline
&&\multicolumn{4}{|c|}{segmentation}& \multicolumn{2}{c}{PSA}   \\
&&&&9 planes&11 planes&&\\
&&&&$\&$&$\&$&&\\
&full crystal&9 planes&11 planes&6 sectors&6 sectors&3 mm&5 mm\\
\hline
2 x 1020 keV&94.7&86.4&84.8&83.7&82.4&86.8&88.7\\
1500 + 540 keV&93.9&84.8&83.1&82.0&80.5&85.6&87.5\\
1734 + 306 keV&92.9&82.9&81.0&79.8&78.2&83.9&85.8\\
2040 keV&90.9&80.1&78.3&77.0&75.5&80.4&82.5\\
 \hline
 single electron & 93.8 & 84.6 & 82.9 &
 81.7 & 80.3 & 85.4 & 87.3 \\
 $0\nu\beta\beta$ spectrum  & &&&&&& \\ \hline

\end{tabular}
\label{DBD-4kg}
\end{center}
\end{table*}

\subsection{Sensitivity}

The overall detection efficiency to signal estimated above has
been used together with Eqs. \ref{detectorfom} and \ref{mef} to
compare the sensitivity of germanium DBD experiments using
different background rejection schemes. Three different situations
have been taken into consideration: no rejection technique, the
application of anticoincidence rejection for crystals with
6$\times$11 segments, and the use of PSA techniques assuming a 3
mm spatial resolution. Detectors enriched in $^{76}$Ge at 86\% and
total exposures of MT=100 and 1000 kg$\cdot$y have been assumed.
An energy window of 3.5 keV has been considered, as in Ref.
\cite{avignonenjp}. Values of background level and corresponding
detection efficiency deduced in this work for 4-kg crystals have
been used for each situation. Table \ref{sensitivities} summarizes
the evaluated sensitivities, presenting the F$_{D}$ values and the
corresponding effective neutrino masses considering the average
nuclear factor-of-merit F$_{N}$=7.3$\times$10$^{-14}$ y$^{-1}$
used in Ref. \cite{avignonenjp}. Neutrino masses below 50 meV can
be explored with very segmented crystals or applying PSA
techniques and for high enough exposures.

\begin{table*}
\begin{center}
\caption{Comparison of the sensitivity of germanium DBD
experiments under different background rejection schemes,
evaluated following Eqs. (\ref{detectorfom}) and (\ref{mef}) (see
text).} \centering
\begin{tabular}{l|ccccc}
\hline

& MT (kg$\cdot$y) & b (c keV$^{-1}$ kg$^{-1}$y$^{-1}$) &
$\epsilon$ (\%) & F$_{D}$ (10$^{26}$ y) & $<$m$_{\nu}$$>$ (meV) \\
\hline

no rejection & 100 & 0.022 & 93.8 & 1.6 & 149 \\
6$\times$11 segmentation & 100 & 0.0019 & 80.3 & 4.7 & 88 \\
PSA (3 mm resolution) & 100 & 0.0011 & 85.4 & 6.5 & 74 \\ \hline
no rejection & 1000 & 0.022 & 93.8 & 5.1 & 84 \\
6$\times$11 segmentation & 1000 & 0.0019 & 80.3 & 15 & 49 \\
PSA (3 mm resolution) & 1000 & 0.0011 & 85.4 & 21 & 42 \\
\hline
\end{tabular} \label{sensitivities}
\end{center}
\end{table*}

\begin{figure}
\centering
  \includegraphics[height=.2\textheight]{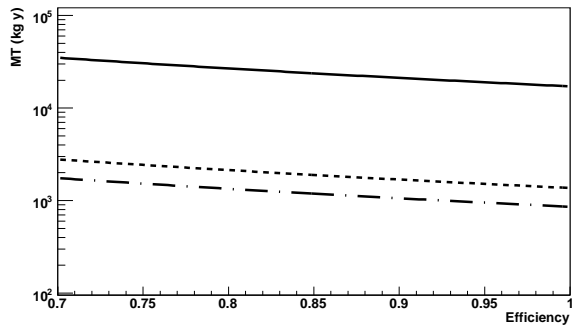}
  \caption{Exposure MT (in kg$\cdot$y) necessary to reach the sensitivity to explore neutrino effective masses $<m_{\nu}>$ of 40 meV,
 versus the detection efficiency to
  $0\nu\beta\beta$ events considering full crystals (solid line), anticoincidences in 11 x 6 segmented crystals (dashed line), and PSA with a spatial resolution for PSA of 3 mm
  (dot and dash line). }
  \label{MT_Eff}
\end{figure}

To achieve a certain sensitivity to the effective neutrino mass,
the required exposure of an experiment depends on the background
level in the region of interest and the detection efficiency to
the signal (see Eqs. \ref{detectorfom} and \ref{mef}); both
background level and efficiency are different when different
background rejection schemes are considered. Figure \ref{MT_Eff}
shows this dependency between the exposure and the efficiency in
the three background rejection schemes considered before. Using
the overall signal efficiency values previously estimated, it can
be deduced from this plot that despite the loss of efficiency,
segmented crystals working in anticoincidence require an exposure
one order of magnitude lower than that of non segmented detectors
to explore the same range of neutrino masses. If PSA techniques
are used with 3 mm of energy resolution, an additional factor of
$\sim$two of reduction is achieved in the necessary exposure.

\section{Discussion and conclusions}
\label{conclusions}

Some conclusions can be drawn from the analysis of the different
strategies for background reduction in germanium double beta decay
experiments. The study of the granularity of the detector system
shows that heavier crystals are better to reduce the contribution
of external radioimpurities, but worse to reduce background coming
from internal contaminations. By applying the most powerful
segmentation techniques taken here into consideration in a 4-kg
detector, 2(5) out of 100 events due to internal impurities from
$^{60}$Co ($^{68}$Ge) would remain in the RoI, while for external
contaminations, about half of the events would be rejected. A
3-dimensional spatial resolution of 3 mm, obtained by means of PSA
in segmented detectors, would allow to reject more than 99\% of
background events due to cosmogenic isotopes induced in the
crystal, and around 60\% of those coming from external 2614.5 keV
photons.

According to numbers presented in Table
\ref{reducedcountingrates}, a background level of 1.1 (1.6)
$10^{-3}$ c keV$^{-1}$ kg$^{-1}$y$^{-1}$ due to the studied
background sources could be achieved using 4 (2)-kg crystals when
considering very precise conditions. For the production of
$^{60}$Co ($^{68}$Ge), the exposure time is of 30 (180) days and
the production rate is 5 (1) kg$^{-1}$d$^{-1}$. In the case of
$^{68}$Ge, a cooling time of 180 days has been also taken into
account. For the external 2614.5 keV, an environmental flux of 0.1
cm$^{-2}$ s$^{-1}$ has been assumed and the use of a 40-cm-thick
lead shield considered. For intrinsic $^{232}$Th impurities in the
lead shielding, the activity supposed is 1 $\mu$Bq/kg. In these
optimal conditions, the raw background is reduced by more than one
order of magnitude thanks to a 3 mm spatial resolution achieved by
PSA in segmented detectors. The most relevant contribution is that
of the 2614.5 keV produced in the lead shielding itself by
$^{232}$Th impurities, followed by the one from cosmogenic
$^{68}$Ge induced in the crystal. External 2614.5 keV photons and
cosmogenic $^{60}$Co are almost negligible. The use of large
crystals, having less background for external contaminations than
the small ones, seems more adequate since background rejection by
anticoincidence between segments or by PSA is efficient enough in
the reduction of contributions from internal radioimpurities.

Some comments are in order if considering other less favorable
conditions for the raw background or the rejection strategies:

\begin{itemize}

\item Contribution from $^{60}$Co emissions is in general
negligible thanks to the very good rejection factors attainable
either with PSA in segmented detectors or just anticoincidence
techniques between segments of the detectors.

\item $^{68}$Ge contribution in the scenario described above
considering a production rate of 1 kg$^{-1}$y$^{-1}$ is very small.
However, it must be noticed that the production rate calculated in
different estimates presented in Sec. \ref{Production Rates} is much
higher. It has been shown that if the production rate was 10 times
bigger, a cooling time of 2 years instead of 3 months would be
necessary to achieve background contributions of the same order of
magnitude.

\item To shield external 2614.5 keV photons from environmental
gamma radiation in the laboratory, 40 cm of lead are mandatory.
When using just 30 cm, the goal of a background level of $10^{-3}$
c keV$^{-1}$ kg$^{-1}$y$^{-1}$ cannot be fulfilled.

\item Regarding intrinsic $^{232}$Th impurities in the lead
shield, it is worth noting that the achievement of a radiopurity of
1 $\mu$Bq/kg cannot be taken for granted since common upper limits
to contaminations of this chain achieved in lead are of some
hundreds of $\mu$Bq/kg (see for instance the ILIAS database on
radiopurity measurements\footnote{Available at
http://radiopurity.in2p3.fr/. and developed within the ILIAS
(Integrated Large Infrastructures for Astroparticle Science) EU
project.} and even the best limits obtained with the most modern
germanium spectrometers are of some tens of $\mu$Bq/kg
\cite{heusserlrt,laubenstein}. Therefore, to further reduce the
background level from this source, improvement of the radiopurity of
this material or using radiopure copper instead of lead in the inner
part of the shielding would be very useful. It has also been proved
that increasing the thickness of the lead layer is not relevant for
these shielding radioimpurities.

\item For the spatial resolution, if a worse performance of 5 mm
was achieved, the best total background level would be of 1.5
(2.0) $10^{-3}$ c keV$^{-1}$ kg$^{-1}$d$^{-1}$ for 4 (2)-kg
crystals. No dramatic difference in the background level from
2614.5 keV would be produced, while an increase in a factor of
$\sim$3 (4) would be registered for $^{68}$Ge ($^{60}$Co)
emissions, which, fortunately, seem not to be the dominant
background source.

\item The use of 4-kg crystals gives a more reduced background,
but 2-kg detectors, in use for a long time, could be acceptable.
Segmentation schemes assumed are feasible for present germanium
detector technologies and give a best total background level of
1.9 (2.2) $10^{-3}$ c keV$^{-1}$ kg$^{-1}$y$^{-1}$ for 4 (2)-kg
crystals. Therefore, the option of applying just segmentation
techniques must not be discarded.
\end{itemize}

The overall detection efficiency to neutrinoless DBD signals has
been evaluated by Monte Carlo simulation for the different
background rejection scenarios, finding for 4-kg crystals a
reduction from $\sim$94\% to $\sim$80\% when considering
anticoincidences in 11$\times$6 segments and to $\sim$85\% if PSA
techniques are applied with a 3 mm spatial resolution.

The sensitivity of DBD experiments depends on both achieved
background level in the RoI and detection efficiency; it has been
shown that the improvement in the former thanks to rejection
techniques largely compensate the loss in the latter since
experiments with these techniques require much lower exposure for
a fixed sensitivity.

In summary, it seems that contribution from dominant background
sources in previous germanium double beta decay experiments could
be reduced down to $10^{-3}$ c keV$^{-1}$ kg$^{-1}$y$^{-1}$ in the
RoI using present detector technologies, which allows to explore
effective neutrino masses even below 50 meV.

\section{Acknowledgments}

This work has been funded by Spanish MEC contract FPA2004-0974 and
the ILIAS (``Integrated Large Infrastructures for Astroparticle
Science'') EU project (contract number: EU-RII3-CT2003-506222).
Useful discussions within the IDEA (``Integrated Double-beta decay
European Activities'') Joint Research Activity 2 in ILIAS are
deeply acknowledged.



\end{document}